\documentclass[aps,prf,amsmath,amssymb,longbibliography,superscriptaddress]{revtex4-1}
\usepackage[dvipsnames,rgb,dvips]{xcolor}
\usepackage{overpic}
\usepackage{amsmath,amssymb}

\usepackage{graphicx,mathrsfs,psfrag}
\usepackage{dcolumn}
\usepackage{epstopdf}

\usepackage{siunitx}
\usepackage{mathrsfs}

\renewcommand{\cite}{\citep}
\newcommand{\ve}[1]{\ensuremath{\mbox{\boldmath$#1$}}}
\newcommand{\ma}[1]{\ensuremath{\mathbb{#1}}}
\newcommand{\T}{^{\sf T}}
\DeclareMathOperator{\tr}{Tr}

\begin{document}
\title{Lord Kelvin's isotropic helicoid}
\author{Darci Collins}
\author{Rami J. Hamati}
\affiliation{Department of Physics, Wesleyan University, Middletown, CT 06459, USA}
\author{Fabien Candelier}
\affiliation{
Aix Marseille Univ., CNRS, IUSTI, Marseille, France}
\author{Kristian Gustavsson}
\author{Bernhard Mehlig\footnote{\tt bernhard.mehlig@physics.gu.se}}
\affiliation{Department of Physics, Gothenburg University, 41296 Gothenburg, Sweden }
\author{Greg A. Voth\footnote{\tt gvoth@wesleyan.edu}}
\affiliation{Department of Physics, Wesleyan University, Middletown, CT 06459, USA}

\begin{abstract}
Nearly 150 years ago, Lord Kelvin proposed the isotropic helicoid, a particle with isotropic yet chiral interactions with a fluid, so that translation couples to rotation.  An implementation of his design fabricated with a three-dimensional printer is found experimentally to have no detectable translation-rotation coupling, although the particle point-group symmetry allows this coupling.  We explain these results by demonstrating that in Stokes flow, the chiral coupling of such isotropic helicoids made out of non-chiral vanes is due only to hydrodynamic interactions between these vanes. Therefore it  is small.  In summary, Kelvin's predicted isotropic helicoid exists, but only as a weak breaking of a symmetry of non-interacting vanes in Stokes flow.  

\end{abstract}

\maketitle
\section{Introduction}
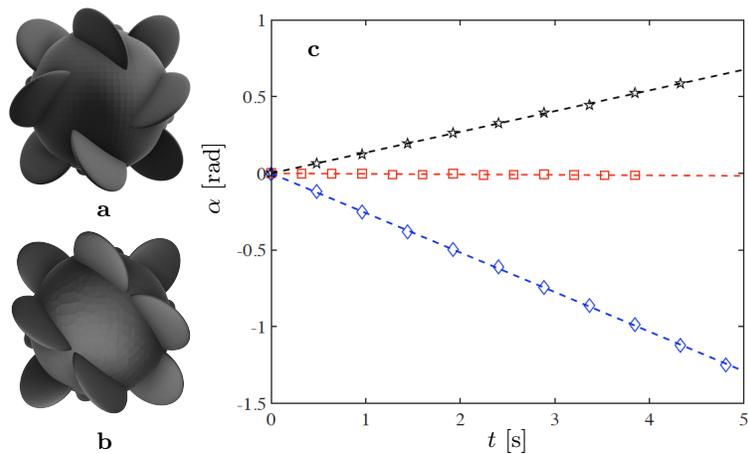
\begin{figure}[b]
 \centering
 \begin{overpic}[width=10cm]{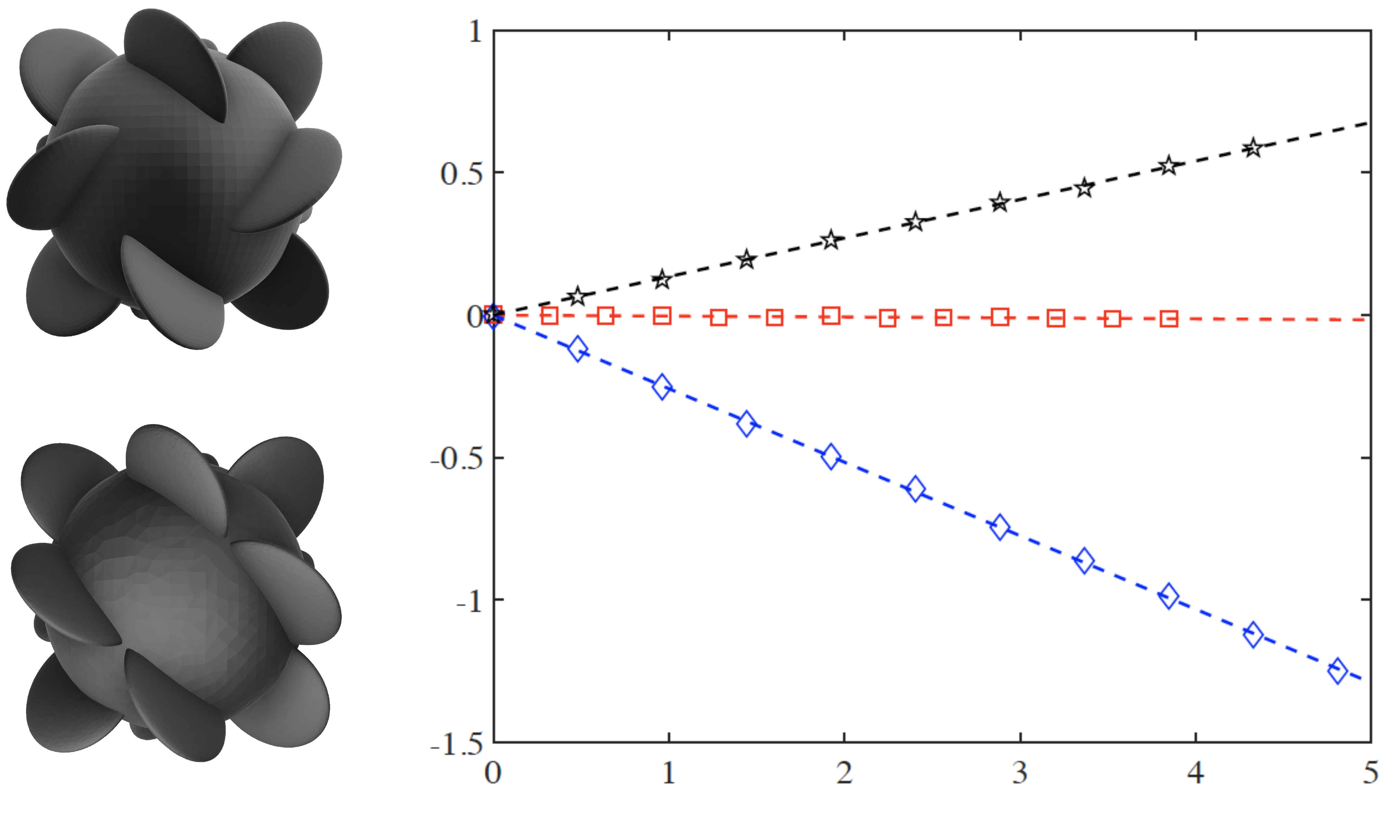}
\put(11,31.7){\colorbox{white}{\bf a}}\put(11,0.7){\colorbox{white}{\bf b}}
\put(39,53){\colorbox{white}{\bf c}}
\put(64,1){$t$ [s]}\put(26,32){\rotatebox{90}{$\alpha$ [rad]}}
\end{overpic}
\caption{\label{fig:exp}({\bf a}) Model used to 3D-print the left-handed isotropic helicoid.  ({\bf b})  Anisotropic helicoid oriented with the equatorial vanes reflected.
({\bf c})  Experimental measurements of the particle  orientation around its sedimentation axis as a function of time. Isotropic helicoid (red $\Box$ with best fit $\omega={\rm d} \alpha/{\rm d}t=-0.003$ rad/s),  anisotropic helicoid with initial orientation as in panel {\bf b} (blue $\Diamond$,  $\omega=-0.258$ rad/s), and with reflected vanes on a meridian (black $\star$,  $\omega=0.135$ rad/s). } 
\end{figure}

 In his analysis of the forces and torques on a rigid body moving in an incompressible inviscid fluid \cite{Kelvin}, Lord Kelvin  
 commented on a particular shape, the {\em isotropic helicoid}, which experiences the same translational resistance in a homogeneous fluid flow at any orientation, just like a sphere. But unlike a sphere, the particle experiences a torque as it moves through the fluid.  To maintain isotropy, this torque must be independent of the particle orientation relative to the flow.
 This may seem surprising if one takes isotropy to mean {\em continuous} rotational symmetry which implies the particle has mirror symmetry and so is non-chiral, which precludes helicity.   
 However Kelvin suggested how to make a helical particle with {\em discrete} rotational symmetry and isotropic drag by placing 12 vanes around the great circles of a sphere~\cite{Kelvin}.  An implementation following his prescription is shown in Fig. \ref{fig:exp}({\bf a}).  
 
 Since Kelvin's analysis of isotropic helicoids in the inviscid limit, textbooks by Happel \& Brenner \cite{Happel:1983} and Kim \& Karrila \cite{Kim:2005} have discussed isotropic helicoids in viscous flows and have concluded that the coupling persists in
 the low Reynolds number limit. 
Chiral interactions in turbulent fluids is  an area of active research~\cite{biferale_2013,Kramel:2016}.   Perturbation theory and numerical simulations have been used to study isotropic helicoids with particle inertia whose translation-rotation coupling causes them to preferentially sample helical regions in viscous flows
that are chaotic~\cite{gustavsson2016preferential} or turbulent \cite{biferale2019helicoidal}.   Quantification of chirality is subtle~\cite{Fowler:2005,Efrati:2014}, 
 and translation-rotation coupling of chiral objects is  an important test case for proposed measures of chirality~\cite{Efrati:2014,Szymczak}.    Coupling of translation of chiral particles to rotation and strain is a promising method for hydrodynamic sorting of 
particles by chirality in viscous flows~\cite{Marcos:2009,Eichhorn:2010,Meinhardt:2012,Marichez:2019}.

The elegant theoretical idea of a helical yet isotropic particle has been in the literature for nearly 150 years, cited as an example illustrating the power of symmetry analysis -- yet there is no published experimental verification of the conjectured  translation-rotation coupling of isotropic helicoids, not in the inviscid limit (Reynolds number Re$\to \infty$), neither in the low-Re limit, nor at any Reynolds number in between.  So we 3D-printed different  isotropic helicoids and measured their translation-rotation coupling while settling through a quiescent fluid.   We tried to achieve as small Re as possible, to avoid
confounding effects due to flow separation at large Re, and because the small-Re limit is easiest to analyze theoretically. 
\mbox{O}ur experiments do not show evidence for translation-rotation coupling.

This is surprising, because our particles had the same discrete symmetries as Lord Kelvin's isotropic helicoid.
Their point-group symmetry allows translation-rotation coupling, and therefore it is expected  to be non-zero in general, provided that there is no other symmetry that forbids this coupling. We explain the experimental null result by showing that there is an additional symmetry that causes the translation-rotation coupling to vanish for helicoids made of non-interacting vanes in the creeping-flow limit. This symmetry is weakly broken by hydrodynamic interactions, which produce a very small coupling, too small measure in our experiments. Our results indicate that the translation-rotation coupling of isotropic helicoids is quite small in general. It remains an open question to engineer isotropic helicoids with maximal coupling.

The remainder of this paper is organized as follows. Section \ref{sec:background} briefly describes the background,
Section \ref{sec:exp} gives details about the experiments that resulted in Fig.~\ref{fig:exp}. The theoretical analysis
is described in Section \ref{sec:theory}, and Section \ref{sec:conclusions} contains conclusions.

\section{Background}
\label{sec:background}
\subsection{Force and Torque in the low-Re limit}
In the low Reynolds number limit, the hydrodynamic force $\ve f$ and torque $\ve \tau$ on an arbitrarily shaped particle in a uniform velocity gradient depend linearly on the slip velocity, the angular slip velocity, and upon the local strain rate
\cite{Happel:1983,Kim:2005}:
\begin{align}
\label{tensors}\begin{bmatrix}\ve f \\\ve \tau\end{bmatrix}
=
\mu \begin{bmatrix}\ma A & {\ma B}^{\sf T} & {\ma G} \\\ma B & \ma C & {\ma H} \end{bmatrix}
\begin{bmatrix}\ve u^\infty - \ve v \\\ve \Omega^\infty-\ve \omega \\\ma S^\infty\end{bmatrix}\,.
\end{align}
Here $\mu$ is the dynamic viscosity of the fluid, $\ve v$ and $\ve \omega$ are the velocity and the angular velocity of the particle,
$\ve u^\infty$ and $\ve\Omega^\infty$ are the undisturbed fluid velocity and half the undisturbed fluid vorticity,
and $\ma S^\infty$ is the strain-rate tensor, the symmetric part of the matrix of velocity gradients of the undisturbed fluid.  Moreover,  $\ma A$ is the drag tensor,  $\ma B$ is the translation-rotation coupling tensor,  $\ma C$ is the rotational drag tensor, and   $\ma G$ and $\ma H$ are third-rank tensors that couple force and torque to the strain rate.  They produce effects such as Jeffery orbits~\cite{Jeffery:1922,einarsson2016a}, rectification of rotations of chiral dipoles~\cite{Kramel:2016}, and separation of particles according to chirality~\cite{Marcos:2009,Eichhorn:2010,Meinhardt:2012,Marichez:2019}.
Here we consider a particle settling steadily in a quiescent fluid, so that $\ve u^\infty =0, \ve \Omega^\infty=0$, and $\ma S^\infty=0$. 
In the context of Eq.~(\ref{tensors}), the question  is whether or not Lord Kelvin's particle has non-zero $\ma B$.

\subsection{Symmetries}
\begin{figure}[b]
\begin{overpic}[width=7.cm]{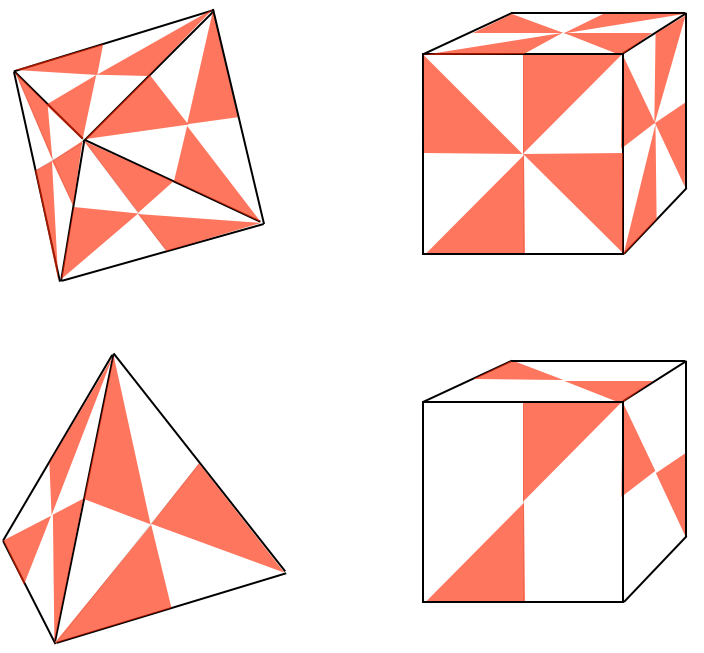}
\put(0,90){{\bf a}}\put(55,90){{\bf b}}\put(0,40){{\bf c}}\put(55,40){{\bf d}}
\end{overpic}
\caption{\label{fig:symmetries}  Illustration of the point-group symmetries of different isotropic helicoids. 
({\bf a},{\bf b})  Chiral octahedral symmetry O,   four C$_3$ rotation axes, three C$_4$ axes, and six C$_2$ axes.
 ({\bf c},{\bf d})  Chiral tetrahedral symmetry T, four C$_3$ rotation axes, and three C$_2$ axes (the cubes were drawn after Table~7.2 in Ref.~\cite{Ashcroft}). }
\end{figure}
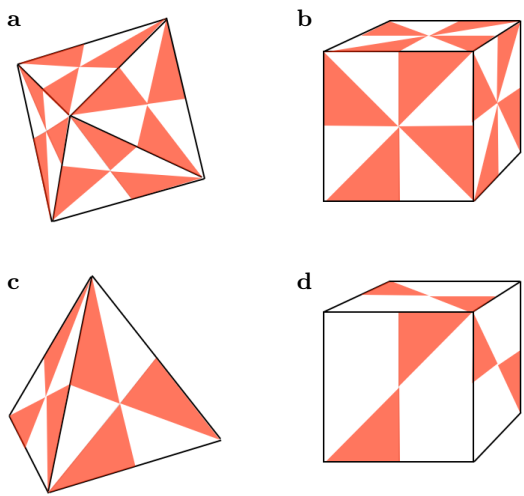
Fig.~\ref{fig:symmetries} illustrates the point-group symmetries of different isotropic helicoids.
  Kelvin's isotropic helicoid places oblique vanes on the 12 edges of an octahedron and has chiral octahedral symmetry, point group O [Fig.~\ref{fig:symmetries}({\bf a})]. 
We also fabricated an isotropic helicoid with six four-armed propellers on the faces of a cube (Section \ref{sec:exp} and 
Appendix~\ref{app:largere}). This particle looks like Fig.~\ref{fig:symmetries}({\bf b}), it also has  chiral octahedral symmetry.
  Panels~({\bf c}) and ({\bf d}) illustrate isotropic helicoids with chiral tetrahedral symmetry, point-group T.
Either point group, O or T, constrains $\ma A$ and $\ma C$ to be proportional to the unit matrix.    Since the groups O and T contain rotations only, 
and no mirror symmetries~\cite{Curtis,Weyl}, the tensor $\ma B$ is constrained in the same way as $\ma A$ and $\ma C$.
Therefore it must be proportional to the unit matrix.    
We conclude that the point-group symmetry of these particles ensures isotropy and allows chiral coupling.
More detailed calculations arriving at the same conclusion are found in Refs.~\cite{Happel:1983,Kim:2005}.  For convenience, we give the details
in Appendix \ref{app:pgs}, for an isotropic helicoid with chiral tetrahedral symmetry.
We note that particles with full tetrahedral or octahedral symmetry with reflections
constrain the translation-rotation coupling to zero, $\ma B=0$.

\section{Experiments}
\label{sec:exp}
We fabricated 
Lord Kelvin's isotropic helicoid [Fig.~\ref{fig:exp}({\bf a})] using a Form 2 stereolithography 3D printer. For vanes, we used 
spheroidal disks with diameter 8.7 mm and aspect ratio 0.2 projecting from the sphere as shown in Fig. \ref{fig:exp}({\bf a}).  As proposed by Lord Kelvin, the centers of the disks are equally spaced around the three great circles of a solid sphere with 45$^\circ$ inclination angle in the direction to create a left-handed particle (using the convention in Ref.~\cite{gustavsson2016preferential}) with chiral 
octahedral symmetry [Fig.~\ref{fig:symmetries}({\bf a})].
 The diameter of the sphere is $d=17.4$ mm and the 3D printed particle has density  $\varrho_{\rm p}= 1.16\,$g/cm$^3$.  When dropped in silicon oil with kinematic viscosity $\nu=5$ cm$^2$/s and density $\varrho_{\rm f}= 0.98\,$g/cm$^3$ in a tank of dimensions 21 cm wide, 10.5 cm thick and 30 cm tall, the particle settles at velocity $v_s=4.74\,$cm/s.
This corresponds to Reynolds number ${\rm Re} = d v_s/\nu =1.65$.

Figure~\ref{fig:exp}({\bf c}) shows the orientation defined as the angle of the particle around the sedimentation axis as it settles approximately $20\,$cm through the chamber.  Here positive $\alpha$ is the rotation direction favored by a left-handed propeller which is clockwise when viewed from above.   The red squares show essentially no change in orientation for the isotropic helicoid with a best fit angular velocity of $\omega=-0.003$ rad/s.    
For comparison, we also printed the anisotropic helicoid shown in Fig. \ref{fig:exp}({\bf b}) with the same dimensions but with the four vanes along one of the great circles reflected to form a right-handed propeller. 
Separate experiments were performed with the particle oriented along the principal axes of $\ma B$.  When the right-handed great circle is on the equator so that
its  normal  points in the direction of gravity, the particle rotation is shown in blue diamonds.  When the right-handed great circle is on a meridian, particle orientation is shown in black triangles.   In both cases the sedimentation velocity is within 1\% of that of the isotropic helicoid.  Interestingly, the right-handed rotation rate of the anisotropic helicoid is almost exactly twice the two orthogonal left-handed rotation rates, so for randomly oriented particles the total particle helicity would be nearly zero.  Experimentally we find that the rotation rate of the isotropic helicoid is near zero, only 1.1\% of the maximum rotation rate of the anisotropic helicoid.   The small measured vertical component of the isotropic helicoid rotation rate is twice the random uncertainty in the measurement from video frames, but we observe similar small rotation rates in the other components which are not allowed for homogeneous bodies at low Reynolds numbers and note that density inhomogeneities in the 3D printing are a likely cause.  We conclude it is a null detection with approximately 1\% uncertainty.

We  tried  other versions of isotropic helicoids, but none of them exhibited translation-rotation coupling within measurement error.   Four helices extending along tetrahedral angles showed very little rotation but suffered from fabrication imperfections.  An isotropic  helicoid of six model-airplane propellers in a cubic configuration 
(details in Appendix~\ref{app:largere}) was used in some exploratory experiments on rotation-translation coupling in the high Reynolds-number regime. 
This particle also has chiral octahedral symmetry [Fig.~\ref{fig:symmetries}({\bf b})]. Its four-bladed propellers (Hobbyking hy 10X8.25) formed a cube with sides 18 cm which when dropped in air over a distance of 5.5 m reached a Reynolds number of more than $10^5$, but no significant reproducible rotation was observed in several different drop orientations.  A single propeller dropped with the blades in the horizontal plane rotated at 46 rad/s after falling the the same distance in 2.5 seconds.   This high Reynolds case where Eq.~(\ref{tensors}) does not apply is more complicated since there can be separation and symmetry breaking instabilities which can produce tumbling trajectories that couple translation and rotation even for non-chiral particles~\cite{ern:2012}.
Therefore we  focus  on the low Reynolds number limit in the remainder of this paper.

\section{Theory}
\label{sec:theory}
Our experiments showed no measurable translation-rotation coupling, yet symmetry analysis allows this coupling, as described
in Section \ref{sec:background}. One might argue that the coupling constant could vanish for Lord Kelvin's particle because the vanes around the equator contribute strongly to the expected helical coupling, yet the remaining eight vanes create a torque in the opposite sense, potentially cancelling the coupling.  This opposite torque may be familiar to anyone who has observed a  propeller move through fluid perpendicular to its usual motion, rotating with helicity opposite to the one usually considered~\cite{Efrati:2014}. 

This is unlikely, however, because of the general principle that says: if symmetry allows a coupling to be non-zero, 
then this coupling does not vanish unless there is another, undetected symmetry that constrains the coupling to zero. 
We now demonstrate that there is in fact such an additional symmetry for isotropic helicoids  consisting of vanes that do not interact hydrodynamically with each other. In the low-Re limit, hydrodynamic interactions break this additional symmetry only weakly  for the isotropic helicoids we considered. This explains why our particles have very small translation-rotation couplings.

\subsection{Independent vanes}
An isotropic helicoid made out of non-interacting, non-chiral vanes has zero translation-rotation coupling $\ma B$.   To prove this, start with the  translation theorem~\cite{Happel:1983,Kim:2005} which allows us to relate the tensor $\ma B$ for a helicoid made out of $M$ non-interacting vanes
to the resistance tensors of its vanes, $\ma A^{\!({\rm v})}$ and $\ma B^{({\rm v})}$:
\begin{align}
\label{eq:tt}
\ma B= \sum_{m=1}^{M}\big[ \ma O_m
\ma B^{({\rm v})}  \ma O_m^{\sf T} + \ve r_{\!m} \wedge {(}\ma O_m \ma A^{\!({\rm v})} \ma O_m^{\sf T}{)}\big]\,.
\end{align}   
Here $\ma{O}_m$ is the rotation matrix that rotates from the eigenframe of the isotropic helicoid to that of vane $m$,
$\ve r_m$ is the translation vector from the 
origin of the particle to the centre of the vane,
and the matrix $\ve v \wedge \ma W$ 
has elements $\varepsilon_{ijk} v_j W_{k\ell}$, summation over repeated indices implied. 
Now a non-chiral vane has $\ma B^{({\rm v})}=0$ because of mirror symmetry, and the translational resistance tensor is always symmetric \cite{Happel:1983}, 
$[\ma A^{\!({\rm v})}]^{\sf T} = \ma A^{\!({\rm v})}$. Using the antisymmetry of the vector product in the second term on the right hand side  of Equation (\ref{eq:tt}) it follows
that $\tr \ma B=0$~\footnote{There appears to be a counter example to this theorem in an example in Ref.~\cite{Kim:2005} on page 120 where the translation rotation coupling for a propeller made of unequal size disks has non-zero trace.  But it turns out that this is a typographical error with a non-diagonal element printed on the diagonal.}.    Since $\ma B$ must be proportional to the unit matrix for an isotropic helicoid, it follows that $\ma B$ is identically zero.
A similar result for the rotational coupling of filaments due to collisional momentum transfer is derived in Ref.~\cite{Szymczak}.     
A consequence is that any helicoid made from non-interacting non-chiral vanes must have zero mean rotation when averaged over all orientations as observed in the experiments.  

\subsection{Hydrodynamic interactions}
However, the vanes in a helicoid have hydrodynamic interactions between them.  We now show that these interactions produce non-zero chiral coupling as allowed by symmetry, but that the coupling is quite small because the contribution from independent vanes vanishes.
 To obtain the leading-order hydrodynamic corrections we  determine how a given vane $n$ is affected by the disturbance flow created by the other vanes, assuming that the latter are independent. This corresponds to the first-order terms obtained 
 in a systematic expansion in $b/r$ where $b$ is the vane size, and $r$ is the separation between two neighbouring vanes
 \cite{Kim:2005}.
 
 Consider an  isotropic helicoid made out of $M$ non-chiral vanes. Each vane $m$ has zero translation-rotation coupling, 
 $\ma{B}_m^{({\rm v})} = 0$, and its drag tensor in the frame of the helicoid is denoted by  $\ma{A}_m^{({\rm v})}=\ma O_m \ma{A}^{({\rm v})}
 \ma O_m^{\sf T}$, where $\ma O_m$ is the appropriate rotation matrix, as defined above.
 When a single vane moves in a fluid at rest at position $\ve {x}_{\rm  v}$ with velocity $\ve {v}_{\rm  v}$, it produces the disturbance flow 
\begin{equation}
\ve {u}_{\rm v}' {(\ve x)}= \frac{1}{8 \pi \mu}  \ma{J} {(\ve x-\ve x_{\rm v})} \ve {f}_{\rm v} + \mathscr{O}\Big(\frac{1}{|\ve x-\ve x_{\rm v}|^2}\Big) \
\end{equation}
at position $\ve x$, with $\ve {f}_{\rm v} = \ma{A}^{({\rm v})}   \ve {v}_{\rm v}$.
Here $({1}/{8\pi \mu}) \ma{J}$ is the Green-tensor of the Stokes equation, Equation (\ref{eq:fv}) in Appendix \ref{app:reflection},
and we follow the standard convention, denoting the disturbance flow with a prime.

Each vane moves in the disturbance flow produced by the other vanes. To lowest order
in the reflection method, we approximate the force that  vane $m$ exerts  upon the fluid by
\begin{equation}
\label{eq:fv}
\ve f_{{\rm v}, m} = \ma{A}_m^{({\rm v})} \Big( 
\ma{I}
 - \frac{1}{8\pi \mu}  \sum_{m^\prime \neq m }^M
 \ma J (\ve x_m - \ve x_{m^\prime} ) \ma A^{({\rm v})}_{m^\prime} 
 \Big)  \ve {v}_{\rm v}\,,
\end{equation}
where $\ma{I}$ is the identity tensor, $\ma{A}_m^{({\rm v})} \equiv \ma{O}_m \ma{A}^{({\rm v})} \ma{O}_m^{\sf T}$,
and $\ve x_m$ is the position of vane $m$.
Eq.~(\ref{tensors}) shows that the drag tensor of vane $m$ in the presence of the other vanes is given by 
$ \ma{A}_m^{({\rm v})} + \delta  \ma{A}_m^{({\rm v})}$ with
\begin{equation}
\label{eq:r10}
 \delta \ma{A}_m^{({\rm v})} =  - \frac{1}{8\pi \mu}  \sum_{m^\prime \neq m }^M
\ma{A}_m^{({\rm v})}   \ma J (\ve x_m - \ve x_{m^\prime} )  \ma A^{({\rm v})}_{m^\prime} \,.
 \end{equation}
 At leading order in $|\ve x_m-\ve x_{m'}|^{-1}$,  the drag and translation-rotation coupling  tensors of the isotropic helicoid read
  \begin{equation}
  \label{eq:Bp}
 \ma{A}= \!\!\sum_{m=1 }^M  \ma{A}_m^{({\rm v})} + \delta \ma{A}_m^{({\rm v})}  \, \,\mbox{and} \,
 \,\ma{B}  = \!\!\!\sum_{m=1}^M \ve r_m \wedge (\ma{A}_m^{({\rm v})} + \delta \ma{A}_m^{({\rm v})}) \,.
 \end{equation}
The drag tensor  $ \ma{A}$ must be symmetric.   Eq.~(\ref{eq:Bp}) is consistent with this requirement because $\sum_{m=1}^M  \delta \ma{A}_m^{({\rm v})} $ is symmetric even though the 
 individual $\delta \ma{A}_m^{({\rm v})}$
need not be symmetric. As a consequence we find that
$
 \tr \ma{B}   = \tr ( \sum_{m=1}^M\ve r_m \wedge  \delta  \ma A^{({\rm v})}_m )\neq 0
$
in general. In other words, hydrodynamic interactions between the vanes cause non-zero translation-rotation coupling for an isotropic helicoid.

 One might expect that the trace of  $ {\ma B}$ tends to zero as the size of the particle tends to infinity (keeping the vanes unchanged), because
 hydrodynamic interactions in the  low-Re limit are negligible between distant objects. However this argument fails for the translation-rotation coupling.
 It is true that hydrodynamic interactions decay as $|\delta\ma A^{({\rm v})}_m|\sim  |\ve r_m|^{-1}$, 
 because $\ma J$ decays in this way. But this decay is cancelled by  the magnitude of $\ve r_m$ in the vector product
in Eq.~(\ref{eq:Bp}), so that $\tr {\ma B}$ tends to a constant as $|\ve r_m|\to\infty$. This means that hydrodynamic interactions between the vanes of an isotropic helicoid are not negligible, even if the vanes are very far apart from each other. An explicit calculation for an example is given in Appendix \ref{app:example}.
 We mention that at non-zero Re, the Stokes solution breaks down at large distances, resulting in additional corrections
 to $\tr {\ma B}$.
 
\subsection{Numerical simulations}
To test the theory, Eqs.~(\ref{eq:r10}) and (\ref{eq:Bp}), we computed higher-order hydrodynamic corrections 
 using the method of Durlofsky {\em et al.} \cite{durlofsky1987dynamic}, a variation of the method of reflections \cite{Kim:2005}. 
We considered an isotropic helicoid made out of 24 spheres of radius $a$  linked by massless rigid rods [Fig. \ref{fig:theory}({\bf a})],
with the same point-group symmetry as the particle in Fig.~\ref{fig:exp}({\bf a}). Each vane is modelled as a dumbbell {(length $b=5a$) consisting of two spheres.
Each dumbbell is tangential to the surface of an imaginary sphere of radius $c$ (the radius of the isotropic helicoid), 
and inclined at 45$^\circ$, just like the vanes in Fig.~\ref{fig:exp}({\bf a}).
Details of the method are described in Appendix~\ref{app:reflection}.  

Fig.~\ref{fig:theory}({\bf b}) shows our numerical results for the magnitude of the translation-rotation coupling as a function of particle size, $c$. 
Our numerical results are shown as a solid line, the first-order theory~[Eqs.~(\ref{eq:r10}) and (\ref{eq:Bp})] as a dashed line. 
We see that the numerical results approach the theory [Eqs.~(\ref{eq:r10}) and (\ref{eq:Bp})] at large values of $c$. This is expected, because the contribution of the first reflection must dominate when the vanes are far apart. The convergence is
very slow however, the difference between numerical results and lowest-order theory scales as $c^{-1}$ (not shown).
 
\begin{figure}[t]
\raisebox{2mm}{\begin{overpic}[width=5cm]{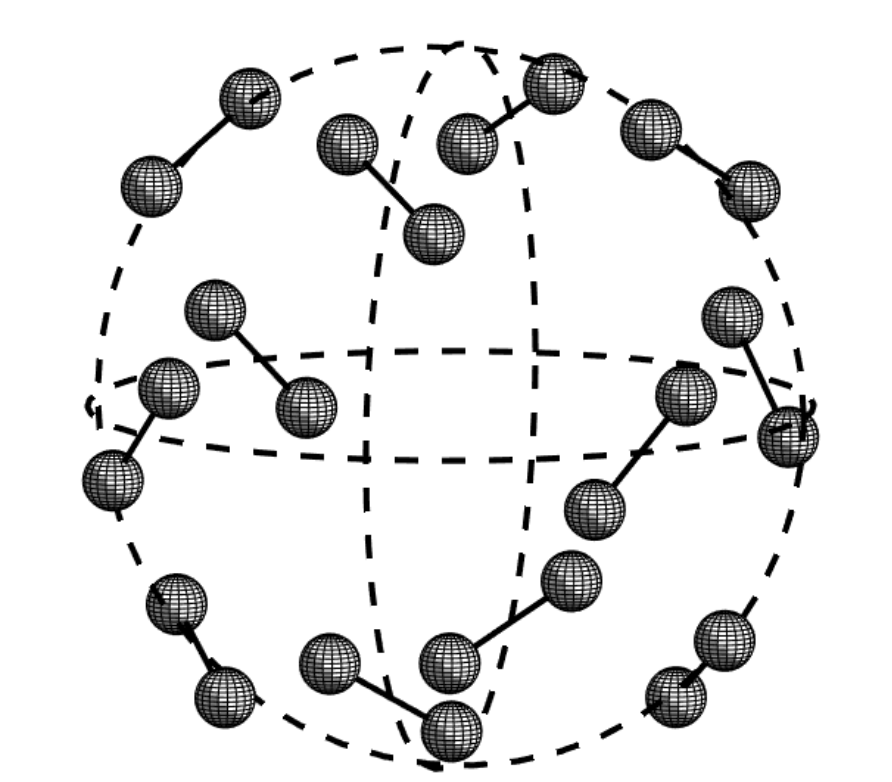}
\put(2,12){{\bf a}}
\end{overpic}}\hspace*{8mm}
\begin{overpic}[width=5.75cm]{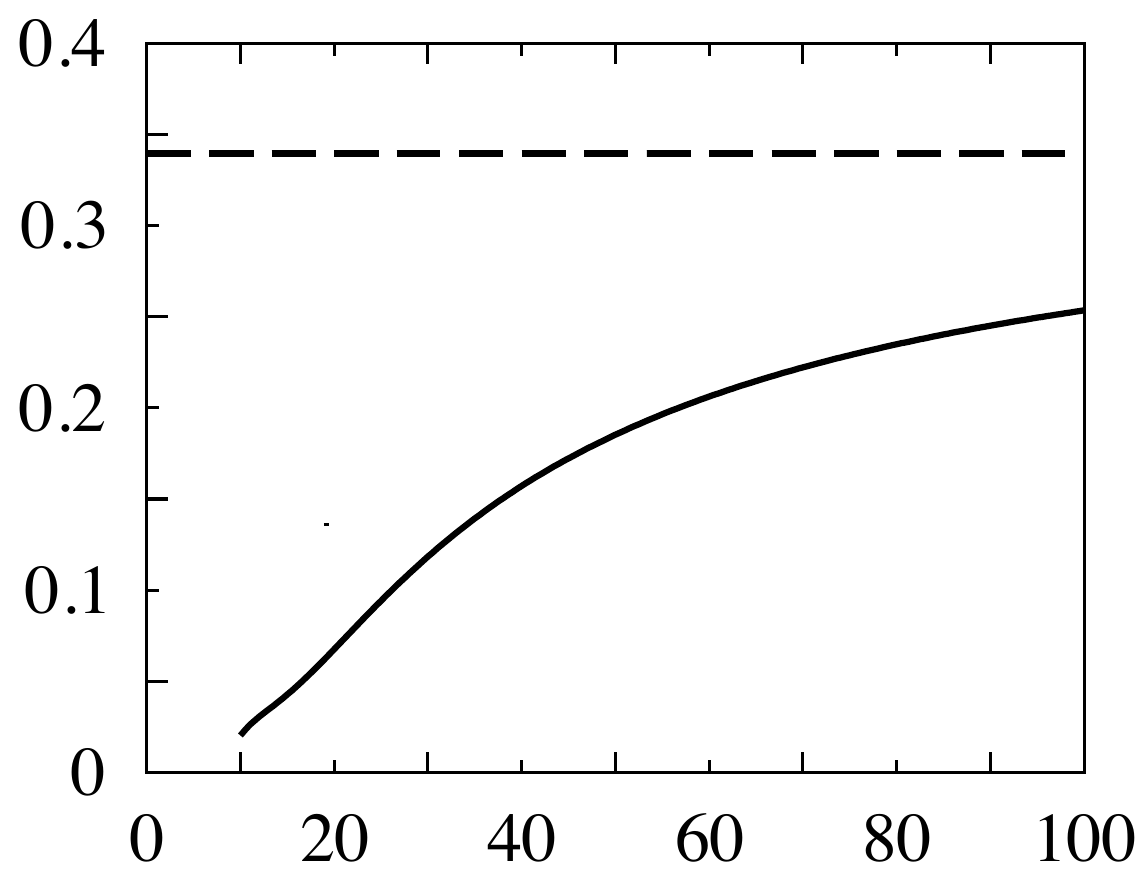}
\put(80,16){\colorbox{white}{{\bf b}}}
\put(-24,62){${\tr \ma B}/{a^2}$}
\put(63,-3){$c/a$}
\end{overpic}
\caption{\label{fig:theory} Theory. ({\bf a})  Schematic of an isotropic helicoid made out of 24 spheres (radius $a$) arranged into 
 twelve dumbbells. Each dumbbell (distance $b=5a$ between centres of spheres)  represents a vane of the particle
shown in Fig.~\ref{fig:exp}({\bf a}). The dumbbells are assumed to be rigidly connected to each other. Three dashed great circles (radius $c$) are guides to the eye.
({\bf b}) Trace $\tr \ma B$ of the translation-rotation coupling matrix for the isotropic helicoid  from panel {\bf a}, as a function
of particle size $c/a$. Numerical result (solid line), Eqs.~(\ref{eq:r10}) and~(\ref{eq:Bp}) (dashed line).
}
\end{figure}

To compare with the experiments, we computed the
steady-state angular velocity for the isotropic helicoid in Fig.~\ref{fig:theory}({\bf a)}. To this end one starts from the expression
for the total force $\ve F$ and torque $\ve T$ on a settling particle in a quiescent fluid in the low-Re limit:
\begin{align}
\label{tensors1}
\begin{bmatrix}\ve F \\
\ve T\end{bmatrix}
=
-\mu \begin{bmatrix}\ma A & {\ma B}^{\sf T}  \\
\ma B & \ma C  \end{bmatrix}
\begin{bmatrix}\ve v \\
 \ve \omega \end{bmatrix}+
(m_{\rm p}-m_{\rm f})\begin{bmatrix} \ve g\\0\end{bmatrix}\,.
\end{align}
The first term represents the hydrodynamic force and torque in the low-Re limit including
hydrodynamic interactions, and the second term is due to gravity, with
gravitational acceleration $\ve g$ with particle mass $m_{\rm p}$, and fluid mass $m_{\rm f}$.
To obtain the steady-state angular velocity we must set $\ve F$  and $\ve T$ to zero. The resistance tensor can be inverted using
standard formulae \cite{22inversion} for the inversion of block matrices:
\begin{equation}
\label{eq:omega}
\ve \omega = -\ma C^{-1}\ma B \ve v \quad\mbox{with settling velocity}\quad \ve v = (m_{\rm p}-m_{\rm f})\mu^{-1} (\ma A-\ma B^{\sf T}\ma C^{-1}\ma B)^{-1}\ve g\,.
\end{equation}
We also evaluated the angular velocity of an anisotropic helicoid similar to Fig.~\ref{fig:theory}({\bf a)}, but with the
equatorial dumbbells flipped, as in the experiment. 
Numerical evaluation of Eq.~(\ref{eq:omega}) for these two particles using the method described
in Appendix \ref{app:reflection} indicates
that the ratio of the angular velocities, isotropic to anisotropic, is non-zero but very small.
It is of the order of $10^{-3}$, just below the experimental accuracy. 

Numerical evaluation of Eq.~(\ref{eq:omega}) also indicates that the magnitude of the translation-rotation
coupling and the resulting sense of rotation of the isotropic helicoid depend sensitively on the precise nature of the hydrodynamic interactions.  A helicoid with a 25th sphere in the centre of the particle}, for example, has the opposite sense of rotation than the particle shown in Fig.~\ref{fig:theory}({\bf a)}.    
     
Suppose we average the angular velocity of the isotropic helicoid over initial orientations. Our numerical results
show that the average angular velocity  is not quite zero, because of hydrodynamic interactions. From Eq.~(\ref{eq:omega}) we find  $\langle \omega_g/v_g\rangle = -\tr \ma C^{-1}\ma B\propto \tr \ma B$ to leading order, 
where $ v_g = \ve v\cdot \hat{\bf g}$, $ \omega_g = \ve \omega \cdot \hat{\bf g}$, and $\hat{\bf g} = \ve g/g$.
Eq.~(\ref{eq:Bp}) shows that $\tr \ma B$  is non-zero, so  that in general $\langle \omega_g/v_g\rangle\neq 0$.
Our numerical calculations show that similarly $\langle \omega_g\rangle\neq 0$.
It could be of interest to explore whether there is an effect analogous to hydrodynamic interactions for the chiral filaments rotating due to momentum transfer from particle collisions~\cite{Szymczak}. A given segment could shield other filament segments from collisions, or give rise to multiple collisions.

\section{Conclusions}
\label{sec:conclusions}
 In conclusion, we measured the dynamics of an isotropic helicoid 
 suggested by Lord Kelvin 150 years ago as it settles in a viscous fluid.  Although symmetry analysis indicates that the particle should start to rotate as it settles, we did not detect any translation-rotation coupling in our experiments. This raises the question whether Lord Kelvin's original argument is flawed.  Analytic calculation of the rotation-translation coupling tensor for non-interacting, non-chiral vanes shows that the coupling is exactly zero.  
But taking into account hydrodynamic interactions between vanes reveals non-zero translation-rotation coupling.
This coupling is quite weak in general, because it is due to hydrodynamic interactions between the vanes of the isotropic helicoid.   The predicted coupling is too weak to be detected in our current generation of experiments.

The possibility of chiral coupling without anisotropy provides an intriguing way to deviate from simple spherical systems,  independent from, and in some ways simpler than the much studied case of spheroids.  Our discovery of the small size of the chiral coupling helps explain why 150 years after Kelvin first introduced the concept, there are no published measurements of isotropic helicoids.   Designing helicoids with optimal chiral coupling provides a challenging focus for future work since it requires designing to control hydrodynamic interaction.

Our results are in keeping with a general rule pertaining to symmetry arguments, seen also in quantum mechanics \cite{merzbacher1998quantum}:  If a symmetry allows a matrix element to be non-zero, then it does not vanish unless constrained by some other symmetry.  Cases where weak symmetry breaking creates almost zero matrix elements provided deep insights in quantum physics, and we suggest that future work to quantify and optimize isotropic helicoids may also be fruitful. 

An important future question is how fluid inertia affects the translation-rotation coupling.  Our theory in the low-Re limit explains the experimental findings at small Re. In principle it is possible to include Re-corrections in the theory, but we see no reason why they should change the qualitative conclusions. On the other hand it is interesting to determine whether small-Re corrections decrease or increase the coupling. Our experiments at large Re did not show any evidence for a coupling, but they are far out of the regime of validity of a small-Re expansion.

\acknowledgments{We thank S. \"Ostlund, D. Koch, Ji Zhang, and P. Szymczak for discussions. We acknowledge  support from NSF grant DMR-
1508575, Army Research Office Grant W911NF-15-1-0205, Vetenskapsr\aa{}det [grant number 2017-3865], Formas [grant number 2014-585],  and by the grant \lq Bottlenecks for particle growth in turbulent aerosols\rq{} from the Knut and Alice Wallenberg Foundation, grant number 2014.0048.}


%

\appendix

\section{Tetrahedral symmetry}
 \label{app:pgs}
 We want to deduce the translation-rotation coupling matrix $\ma B$ for a tetrahedral isotropic helicoid [{Fig.~2({\bf c})}] from its point-group symmetries. The point-group symmetries of a tetrahedron are summarised in many textbooks, see for example  Refs.~\cite{Curtis,Weyl}. 
The procedure is described in Refs.~\cite{Happel:1983,Kim:2005,Fries}. One assumes that the particle is at
rest, $\ve v = 0$ and $\ve\omega = 0$. Now one determines how the hydrodynamic force and torque transform
under an orthogonal transformatio fn $ \ma R$  corresponding to one of the symmetry operations.
 If the orientation of the particle relative to the flow remains invariant,  the hydrodynamic force becomes simply $\ma R\ve f$.
Since the  torque is multiplied by $-1$  under a reflection,
it  transforms as $\det[\ma R]\, \ma R\ve \tau$,
 provided that the orientation of the particle relative to the flow remains unchanged.
 Inserting this into Eq.~(\ref{tensors}), and using
the orthogonality of $\ma R$ one finds that the resistance
tensors must satisfy the constraints 
\begin{align}
\label{tensortransfBC}\ma A &= \ma R \ma A\ma R\T\,, \quad
\ma B =  \det[\ma R]\,\ma R \ma B\ma R\T\,, \quad
\mbox{and}\quad \ma C =  \ma R \ma C\ma R\T\,.
\end{align}
A standard way \cite{tetrahedron} of parametrizing the corners of the tetrahedral particle shown in
Fig.~2({\bf c})  is in terms of the median vectors $\ve c_1=[1,1,1]^{\sf T}, \ve c_2 = [-1,-1,1]^{\sf T}, \ve c_3= [-1,1,-1]^{\sf T}$, and $\ve c_4=[1,-1,-1]^{\sf T}$ in the Cartesian basis
$\hat{\bf x}$, $\hat{\bf y}$, and $\hat{\bf z}$. 
The particle has {\em chiral tetrahedral}  point-group  symmetry \cite{Curtis,Weyl}.
The symmetry group has $12$ elements. Apart from the identity, the group elements are:
\begin{enumerate}
\item[i.] The $\pi$-rotations around the three bimedians
$\tfrac{1}{2}\{\ve c_1+\ve c_4 -(\ve c_2+\ve c_3)\},
\tfrac{1}{2}\{\ve c_1+\ve c_3 -(\ve c_2+\ve c_4)\}$, and $\tfrac{1}{2}\{\ve c_1+\ve c_2 -(\ve c_3+\ve c_4)\}$
(proportional to the three Cartesian coordinate axes). \label{item1}
\item[ii.] Four clockwise rotations around $\ve c_j$ by $\tfrac{2}{3}\pi$. \label{item2}
\item[iii.] Four counter-clockwise rotations around $\ve c_j$ (by $-\tfrac{2}{3}\pi$). \label{item3}
\end{enumerate}
Since the particle in {Fig.~2({\bf c})}
 does not have any mirror symmetries, $\det[\ma R]=1$ for all symmetry operations in Equation (\ref{tensortransfBC}), so
that the symmetries constrain $\ma A$, $\ma B$ and $\ma C$ in the same way. Therefore  the translation-rotation coupling $\ma B$ is constrained to be a multiple of the unit matrix. In general
this coupling is  expected to be non-zero, as concluded in Refs.~\cite{Happel:1983,Kim:2005}.

To determine the forms of $\ma A$, $\ma B$, and $\ma C$ by an explicit calculation, consider first how the $\pi$-rotations (i) around the Cartesian coordinate axes constrain the elements of the resistance tensors $\ma C$ and $\ma B$.
The corresponding rotation matrices are
\begin{equation}
\label{eq:r1}
\ma R_1 =
\begin{bmatrix}
1 & 0 & 0 \\
0 & -1 & 0\\
0 & 0 & -1
\end{bmatrix}\,,\quad
\ma R_2 = \begin{bmatrix}
-1 & 0 & 0 \\
0 & 1 & 0\\
0 & 0 & -1
\end{bmatrix}\,,
\quad\mbox{and}\quad
\ma R_3 = \begin{bmatrix}
-1 & 0 & 0 \\
0 & -1 & 0\\
0 & 0 & 1
\end{bmatrix}\,.
\end{equation}
Inserting these into Equation (\ref{tensortransfBC})
we find that the tensors must be diagonal in the body-fixed basis:
\begin{equation}
\label{eq:diagonal}
\ma A = \begin{bmatrix}
A_{11} & 0 & 0 \\
0 & A_{22} & 0\\
0 & 0 & A_{33}
\end{bmatrix}\,,
\quad
\ma C = \begin{bmatrix}
C_{11} & 0 & 0 \\
0 & C_{22} & 0\\
0 & 0 & C_{33}
\end{bmatrix}\,,
\quad\mbox{and}\quad
\ma B = \begin{bmatrix}
B_{11} & 0 & 0 \\
0 & B_{22} & 0\\
0 & 0 & B_{33}
\end{bmatrix}\,.
\end{equation}
Now consider how the symmetries {(ii)} constrain the tensors.
The four rotation matrices (angle $\tfrac{2}{3}\pi$) are
\begin{equation}
\label{eq:r2}
\ma R_{{\ve c}_1} =
\begin{bmatrix}
0 & 0 & 1 \\
1 & 0 & 0\\
0& 1 & 0
\end{bmatrix}\,,\quad
\ma R_{{\ve c}_2} = \begin{bmatrix}
0 & 0 & -1 \\
1 & 0 & 0\\
0 & -1 & 0
\end{bmatrix}\,,
\quad
\ma R_{{\ve c}_3} = \begin{bmatrix}
0 & 0 & 1 \\
-1 & 0 & 0\\
0 & -1 & 0
\end{bmatrix}\,,\quad\mbox{and}\quad
\ma R_{{\ve c}_4} = \begin{bmatrix}
0 & 0 & -1 \\
-1 & 0 & 0\\
0 & 1 & 0
\end{bmatrix}\,.
\end{equation}
Inserting these into Equation (\ref{tensortransfBC}) and using
that the tensors    $\ma A$, $\ma B$    and $\ma C$ are diagonal [Equation (\ref{eq:diagonal})], we find that any of the symmetries
$\ma R_{{\ve c}_j}$ gives that $A_{11}=A_{22}=A_{33}\equiv A$, $B_{11}=B_{22}=B_{33}\equiv B$, and $C_{11}=C_{22}=C_{33}\equiv C$. Applying
the remaining symmetries  does not constrain the elements further.

If the faces of the tetrahedron are non-chiral (they do not have propellers), the particle  has a higher symmetry, including
in addition $12$ mirror symmetries.  The corresponding
symmetry group is the {\em   tetrahedral}  group. 
Consider for instance 
a reflection in the plane spanned by $\ve c_3$ and $\ve c_4$.
Take     $\hat {\bf e}$   to be the unit vector $\hat {\bf e} = \ve c_3 \wedge \ve c_4/|\ve c_3 \wedge \ve c_4|$, so that
\begin{equation}
\hat{\bf e} = \tfrac{1}{\sqrt{2}}\begin{bmatrix}
1\\
1\\
0\end{bmatrix}\,.
\end{equation}
Then the
reflection matrix is given by $\ma R_{{\tiny\hat{\bf e}}} = \ma I -2\,\hat{\bf e} {\hat{\bf e}}^{\sf T}$,
\begin{equation}
\ma R_{{\tiny\hat{\bf e}}}  =
\begin{bmatrix}
0 & -1 & 0 \\
-1 & 0 & 0\\
0 & 0 & 1
\end{bmatrix}\,.
\end{equation}
Inserting this matrix into Equation (\ref{tensortransfBC})   and using that  $\ma B$    and $\ma C$ are diagonal [Equation (\ref{eq:diagonal})],
we find that $B_{33}=0$, but $C_{33}$ is not constrained. The other reflection matrices correspond to other
ways of distributing  $-1$ and  $1$ in distinct rows and columns, different from Equations (\ref{eq:r1}) and (\ref{eq:r2}).
These other mirror symmetries constrain $\ma B=0$, whereas
$\ma C$ is not constrained further, apart from that it must be proportional to the identity. In summary, without propellers we have (in the body-fixed basis)
\begin{equation}
\label{eq:diagonal2}
\ma A = \begin{bmatrix}
A& 0 & 0 \\
0 & A & 0\\
0 & 0 & A
\end{bmatrix}\,,\quad
\ma C = \begin{bmatrix}
C& 0 & 0 \\
0 & C & 0\\
0 & 0 & C
\end{bmatrix}\, ,
\quad\mbox{and}\quad
\ma B = 0\,.
\end{equation}

\section{Method for calculating hydrodynamic interactions}
\label{app:reflection}

In our numerical simulations, we take into account hydrodynamic interactions using the method of Durlofsky {\em et al.} \cite{durlofsky1987dynamic}. They developed it to determine the evolution of an assembly of free spheres interacting with each other through hydrodynamic interactions. In this appendix we briefly summarise their method.  We do not  include lubrication effects.
Consider a single sphere of radius $a_s$, at position $\ve x_s$, moving with velocity $\ve v_s$ in an ambient flow $\ve u^\infty$. The disturbance flow produced by this sphere can be determined using the method of singularities, by superimposing a set of singularities built from the Green tensor of the Stokes equations and its spatial derivatives. 

Durlofsky {\em et al.} approximate the disturbance flow produced by a sphere  using a stokeslet, a rotlet, and a stresslet, plus correction terms that take  into account the finite size of the sphere. This means that the solution is approximate, but it is accurate enough for our purposes. The disturbance flow {due to sphere $s$} reads:
\begin{equation}
\label{eq:up}
\ve{u}_s'(\ve x)   = \frac{1}{8\pi \mu} \left[\left( 1 +  \tfrac{a_s}{6} \nabla^2\right)\ma{J}  \ve{f}_{\!s}
+ \ma{R}  \boldsymbol{\tau}_{\!s} + \left( 1 +  \tfrac{a_s}{10}\nabla^2\right) \ma{K} : \ma{S}_s\right]\,,
\end{equation}
where 
\begin{align}
\label{eq:JRK}
[\ma{J}]_{ij} = \frac{\delta_{ij}}{r} + \frac{r_i r_j}{r^3}\:,\quad [\ma{R}]_{ij} = \frac{1}{2} \varepsilon_{ijk} \frac{r_k}{r^3} \:, \quad  [\ma{K}]_{ijk} = \frac{1}{2} \left( \partial_k [\ma{J}]_{ij}+\partial_j [\ma{J}]_{ik}\right)\:, \quad \mbox{and} \quad r_i =  [\ve x -  \ve x_{s}]_ i  \,.
\end{align}
{Further,} $\ve{f}_{\!s}$, $\ve \tau_{\!s}$ and $\ma{S}_s$ are the force, torque,  and the stresslet exerted by sphere $s$ upon the fluid. At this stage they are unknown.  
We mention that  $\nabla^2[\ma{R}]_{ij}$ evaluates to zero, given the expression for the components of $\ma{R}$ {in Eq.~(\ref{eq:JRK}).}

Now consider the disturbance flow $\ve u_s'$ at the centre  $\ve x_s$ of sphere $s$ that is produced by all other spheres (except sphere $s$).
This disturbance is  obtained  by summing Eq.~(\ref{eq:up}) over $s'\neq s${:}
\begin{equation}
\ve u'_s= \frac{1}{8\pi \mu} \sum_{s' \neq s}^{N} \Big[\big( 1 +  \tfrac{a_{s'}}{6} \nabla^2\big)\ma{J}(\ve x_s - \ve x_{s'})  \ve{f}_{\!s'}
+ \ma{R}(\ve x_s - \ve x_{s'})  \boldsymbol{\tau}_{\!s'} + \big( 1 +  \tfrac{a_{s'}}{10} \nabla^2\big) \ma{K}(\ve x_s - \ve x_{s'}) : \ma{S}_{s'} \Big]\,.
\end{equation}
Now force, torque, and stresslet acting upon a sphere  moving in a flow can be calculated using the reciprocal theorem \cite{masoud_stone_2019}. Expanding the disturbance velocity $\ve u'_s$ around the centre of sphere $s$ in order to evaluate the integrals in the reciprocal theorem yields the so-called Fax\`en formulae: 
\begin{subequations}
\label{eq:system}
\begin{eqnarray}
\ve v_s - \ve u^\infty & =& \frac{\ve f_{\!s}}{6\pi \mu a_s}  + \Big( 1 +  \frac{a_s}{6} \nabla^2\Big) \ve u'_s\,, \\
\ve \omega_s - \ve {\Omega}^\infty & =&  \frac{\ve \tau_{\!s}}{8\pi \mu a_s^3}  + \frac{1}{2} \ve \nabla \wedge \ve u'_s\,, \\
\ve  - \ma {S}^\infty & =&  - \frac{\ma{S}_s}{\frac{20}{3} \pi \mu a_s^3}  + \frac{1}{2} \Big(1 +  \frac{a_s}{10} \nabla^2\Big) \left[\ve \nabla \otimes \ve u'_s + (\ve \nabla \otimes \ve u'_s)^{\sf T}\right]\,,
\end{eqnarray}
\end{subequations}
where $ \ma{S}^\infty$ is the symmetric part of the velocity gradient of ${\ve u}^\infty$, ${\ve \Omega}^\infty = (1/2) \ve \nabla \wedge \ve u^\infty$, and $\ve v_s$ and $\ve\omega_s$ are translation and angular velocity of sphere $s$.
Eqs.~(\ref{eq:system}), for $s = 1 \ldots N$ constitute a set of linear equations that can be solved for $\ve f_{\!s}$, $\ve \tau_{\!s}$, and $\ma S_s$.
Durlofsky {\em et al.}  \cite{durlofsky1987dynamic} explain that the error of the solution scales as $\mathscr{O}\big((\tfrac{a}{\ell})^{6}\big)$, where $\ell$ is minimal distance between two spheres. 
This method is quite general. It can be applied to any assembly of free spheres. In our problem the spheres are assumed to be parts of a rigid composite particle. 
This simply means that all spheres are constrained to move with translation velocity $\ve v_s = \ve v+ \ve \omega\wedge\ve r_s$, and angular velocity $\ve\omega_s = \ve \omega$, where $\ve v$ and $\ve \omega$ are the centre-of-mass and angular velocity of the composite particle, and $\ve r_s$ parameterizes the location of sphere $s$ w.r.t. the centre of mass of the composite particle. 
To compute the elements of the force $\ve f$
 we impose $\ve \omega = \ve 0$, and $\ve v= [1,\:0,\:0]^{\sf T}$, ${\ve v}= [0,\:1,\:0]^{\sf T}$, ${\ve v}= [0,\:0,\:1]^{\sf T}$. 
 To compute the torque $\ve \tau$ we impose  ${\ve v} = 0$ and $\ve \omega= [1,\:0,\:0]^{\sf T}$, $\ve \omega= [0,\:1,\:0]^{\sf T}$, $\ve  \omega= [0,\:0,\:1]^{\sf T}$. The elements of the coupling tensor are then obtained from Eq.~(\ref{tensors}). In this way  we computed the numerical results shown in Fig.~\ref{fig:theory}. 
 
 \section{Symmetry breaking due to hydrodynamic interactions}
 \label{app:example}
\begin{figure}[t]
\begin{overpic}[width=8cm]{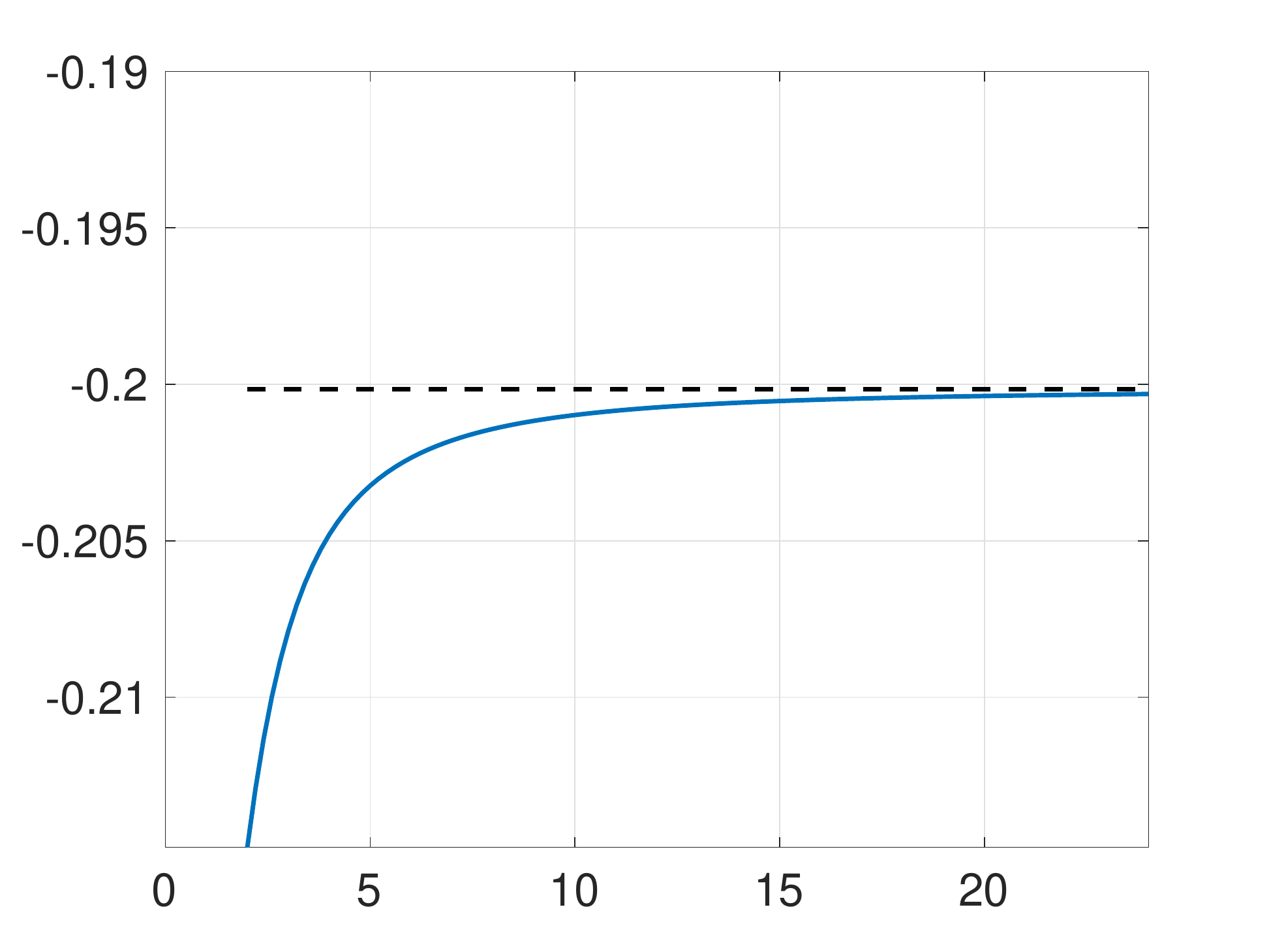}
\put(-10,60){\colorbox{white}{$ \tr \ma{B}^{({\rm p})}$}}\put(61,-3){\colorbox{white}{ $c/b$}}
\end{overpic}
\caption{\label{fig:AppD} Numerical result for trace of $\ma{B}^{({\rm p})}$ of a two-armed propeller made out of two dumbbells,  as a function of the half-distance between the dumbbells, $c$, normalized by $b$ ($a$ equals unity), solid line. Dashed-line shows the theoretical large-$c$ limit, Eq.~(\ref{eq:limBp}).   }
\end{figure}

  To illustrate the effect of hydrodynamic interactions, consider a concrete example,   a propeller  made out of two identical axisymmetric vanes.
 In the body-fixed basis (${\hat{\bf n}}$, ${\hat{\bf  t}}$, ${\hat{\bf  b}}$), the drag tensor of a non-chiral vane reads
 \begin{equation}
  [\ma A^{({\rm v})}]_{i j} = \mu \left[(\mathscr{A}_1+ \mathscr{A}_2)  n_i  n_j + \mathscr{A}_1 ( \delta_{i j}-  n_i n_j )\right]\,,
  \end{equation}
  where the constants $\mathscr{A}_1$ and $\mathscr{A}_2$ parametrise the tensor.
The centres of the vanes are located at $\ve r_1 = c \hat{\bf x}$ and $\ve r_2 = - c \hat{\bf x}$ in the lab frame,
and they are oriented in such a way that their symmetry axes are orthogonal to  ${\hat{\bf x}}$. 
We set $\hat{\bf t}_1 = \hat{\bf x}$  and  $\hat{\bf t}_2 = -\hat{\bf x}$, and we rotate each vane around its own vector ${\hat{\bf t}}$  by an angle $\phi$,
with rotation matrices
\begin{align}
\ma{O}_1 =
\begin{bmatrix}
0 & 1 & 0 \\
\cos\phi & 0& \sin\phi \\
\sin\phi & 0 & -\cos\phi
\end{bmatrix}
\quad \mbox{and}\quad
 \ma{O}_2 = \begin{bmatrix}
0 & -1 & 0 \\
-\cos\phi & 0& -\sin\phi \\
\sin\phi & 0 & -\cos\phi
\end{bmatrix}\,.
\end{align}
Furthermore from Eq.~(\ref{eq:JRK})
\begin{align}
\frac{1}{8 \pi \mu}  \ma{J}(\ve r_2 - \ve r_1) = \frac{1}{8 \pi \mu}  \ma{J}(\ve r_1 - \ve r_2)  = \frac{1}{8 \pi \mu} \frac{1}{2c}
\begin{bmatrix}
2 &0 & 0 \\
0 & 1& 0 \\
0 & 0 & 1 \\
\end{bmatrix}\,.
\end{align}
Using Eq.~(\ref{eq:Bp}) we find for the trace of the translation-rotation coupling tensor 
\begin{align}
\tr \ma{B}^{({\rm p})} = -\frac{1}{8\pi \mu } \tr  \left(  c {\hat{\bf x}} \wedge \ma{O}_1  \ma{A}^{({\rm v})}  \ma{O}_1^{\sf T}    \ma{J}  \ma{O}_2   \ma{A}^{({\rm v})}  \ma{O}_2^{\sf T}  -c {\hat{\bf x}}\wedge \ma{O}_2   \ma{A}^{({\rm v})}  \ma{O}_2^{\sf T}   \ma{J}  \ma{O}_1   \ma{A}^{({\rm v})}  \ma{O}_1^{\sf T} \right)\:.
\end{align}
Since the factors of $c$ in this expression cancel out, the trace of $\ma B^{({\rm p})}$ is independent of $c$ at leading order. This means that the translation-rotation coupling
tensor must tend to a constant in the limit of $c\to\infty$:
\begin{align}
\label{eq:limBp}
\lim_{c\to\infty}\tr \ma B^{({\rm p})}  = \frac{(\mathscr{A}_2)^2 \sin 4\phi}{16 \pi}   + \mathscr{O}\left( \frac{1}{c} \right)\:.
\end{align}
 In order to check this result, we computed the translation-rotation coupling tensor for a propeller made out of two dumbbells using the method described
 in  Appendix \ref{app:reflection}.  The spheres have radius $a$, and the distance $b$ between the spheres that make up the dumbbells is taken to be  $5a$. 
 For this configuration, we find the numerical result  $\mathscr{A}_1 =32.7024 a$ and $\mathscr{A}_2 = -3.4085 a$ for a single dumbbell. Fig.~\ref{fig:AppD} shows  numerical
 results for  $\tr \ma B^{({\rm p})}$ as a function of $c$ for $\phi=\pi/3$.  We see that the numerical results (solid line) approach the theoretical expecation, Eq.~(\ref{eq:limBp}),
 which evaluates to  $\tr \ma B^{({\rm p})}  =  -0.2002 a^2$ for  $\phi=\pi/3$. We add the caveat that the above considerations apply to the Stokes limit. If the Reynolds number 
 is non-zero, the Stokes approximation would fail at large distances, requiring a more elaboration calculation \cite{CandelierMehligDumbbell}. 

\section{Exploratory Experiments at larger Reynolds Number}
\label{app:largere}
\begin{figure}
\includegraphics[width=6in]{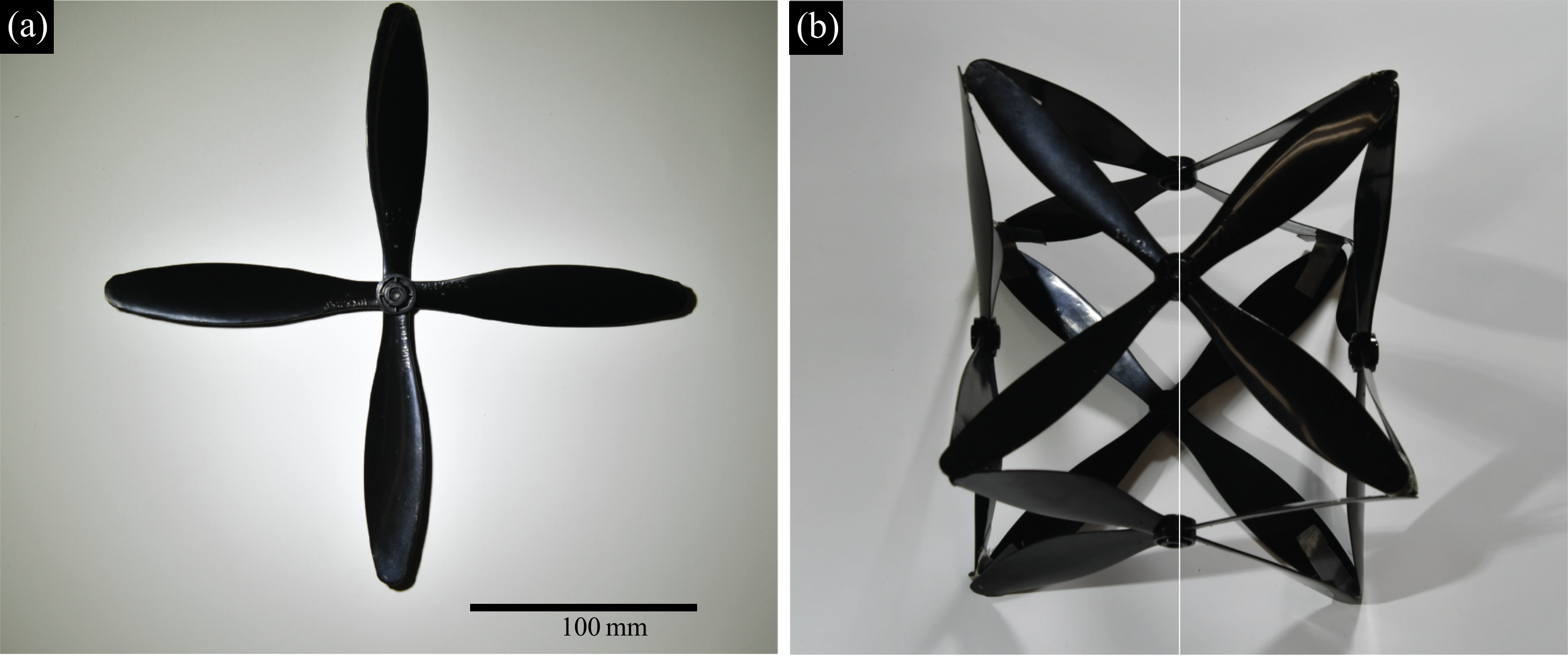}
\caption{{Higher-Re experiments.} ({\bf a}) single four-bladed propeller. ({\bf b}) the cubic isotropic helicoid constructed from six of the propellers shown in panel ({\bf a}).}
\label{fig:prop_cube}
\end{figure}

The four-bladed propeller used in the higher Reynolds-number experiments is shown in Figure~\ref{fig:prop_cube}(a).  It is made by Hobbyking and is an hy10x8.25, which indicates 10 inch diameter and 8.25 inch pitch.   Six of these propellers on the faces of a cube make an isotropic helicoid with chiral octahedral symmetry as shown in Figure~\ref{fig:prop_cube}(b).  Exploratory experiments dropping this helicoid in air over 5.5 m reached a Reynolds number of more than $10^5$, but showed no reproducible translation-rotation coupling suggesting that the observed suppression of translation-rotation coupling for isotropic helicoids extends to high Reynolds numbers.  Further experiments under steady controlled conditions are needed in the high Reynolds number regime.
 
\end{document}